\documentclass[useAMS,usenatbib, usegraphicx]{mn2e}
\usepackage{amsmath}
\usepackage{lscape}

\newcommand{\OIII}{[O~{\sc iii}]}
\newcommand{\OII}{[O~{\sc ii}]}

\newcommand{\NII}{[N~{\sc ii}]}
\newcommand{\SII}{[S~{\sc ii}]}

\newcommand{\HII}{H~{\sc ii}}
\newcommand{\HI}{H~{\sc i}}

\newcommand{\HeII}{He~{\sc ii}}

\title[Ionized gas in Holmberg~II]{Emission spectrum of ionized gas in the Irr galaxy Holmberg II.}

\author[O. V. Egorov et al.]{O. V. Egorov$^1$\thanks{E-mail:
egorov@sai.msu.ru}, T. A. Lozinskaya$^1$\thanks{E-mail:
lozinsk36@mail.ru}, A. V. Moiseev$^2$ \\
$^1$ Lomonosov Moscow State University, Sternberg Astronomical Institute, 13
Universitetskij prospekt, Moscow 119234, Russia\\
$^2$ Special Astrophysical Observatory, Russian Academy of Sciences, Nizhnii Arkhyz 369167, Russia}
\begin{document}

\date{Accepted 2012.... Received 2012...; in original form 2012...}

\pagerange{\pageref{firstpage}--\pageref{lastpage}} \pubyear{2012}

\maketitle

\label{firstpage}

\begin{abstract}

We study the ionized gas spectrum of star forming regions in the Holmberg~II galaxy using the optical long-slit
spectroscopic observations made at the 6-m telescope of the Special Astrophysical Observatory of the Russian
Academy of Sciences (SAO RAS). We estimate the oxygen, nitrogen, sulphur, neon, and argon abundances in
individual \HII\ regions and find the average metallicity in the galaxy to be $Z\simeq0.1$ or $0.3~Z_\odot$
depending on the estimation method employed. 
We use these observations combined with the results of our earlier studies
of the Irr galaxy IC~10 and BCD galaxy VII~Zw~403  to compare the currently most popular 
methods of  gas metallicity estimation in order to select among them the techniques that are
most reliable for analysing Irr galaxies. To this end, we use the `direct' $T_e$ method and
six empirical and theoretical methods. 
The results of our observations mostly confirm the conclusions of
\citet{lopsan} based on the analysis of systematic deviations of metallicity estimates derived by
applying different methods to `model' \HII\ regions.

\end{abstract}


\section{Introduction.}

The irregular galaxy Holmberg II (Ho II, DDO 50, UGC4305, PGC 23324, VII Zw 223)
currently serves as an ideal target for the study of the entire variety of the phenomena
associated with the effect of radiation, stellar winds, and supernovae explosions feedback, which determine the
structure, kinematics, and chemical composition of the interstellar medium as well as
the process of star formation in Irr galaxies.

This gas-rich, non-interacting, and rigidly rotating galaxy is a member of the M81 group
and is located at a distance of  $D=3.39\pm0.20$~Mpc \citep{karach03}. It is an Im
(i.e., Magellanic-type irregular) galaxy with the parameters
$M{_B}=-16.71\pm0.16$~mag, $B-V=0.11\pm0.05$, $E(B-V)=0.032$, and the inclination angle $i=38^\circ$ \citep{moustakas}.

The large-scale structure of the galaxy has been well studied in the optical, radio,
IR, UV, and X-ray bands. A total of about 40 giant cavities and
slowly expanding \HI\ supershells have been found (see \citealt{bagetakos} and
references therein). The stellar population of Ho~II was thoroughly
studied using the \emph{Hubble Space Telescope} (\emph{HST}) data (see, e.g., \citealt{weisz} and \citealt{cook}).

A total of 82 \HII\ regions have been found in the Ho~II galaxy \citep{HSK} and for the
sake of homogeneity we use HSK numbers to identify them throughout this paper.

Bright emission nebulae reside mostly in the `walls' of giant HI supershells and
are concentrated in several `chains' of complexes of ongoing star formation with the
brightest nebulae located in the eastern chain (see Fig.~1 and figures in
\citealt{hunter} and \citealt{karach07}). \citet{stewart} estimate the ages of these
star-forming regions to be between 2.4 and more than 6.3~Myr. The large-scale
distribution of the $\mathrm{H}\alpha$ emission of ionised gas is consistent with that
of FUV emission in the galaxy (see, e.g., fig.~6 of \citealt{stewart} and fig.~29
of \citealt{weisz}). In the eastern chain of the brightest nebulae the brightest
emission of heated dust is observed \citep{walter}.

\begin{table*}
\caption{Log of observations}\label{tab_observ}
\begin{tabular}{crlcccr}
\hline
Spectrum (PA) & Date\,\,\,\,\,\,\,\,\,\,\,\,        &\,\,\,\,\,\,\,\,\, Grism&$\Delta\lambda$, \AA  & $\delta\lambda$, \AA & $T_{exp}$, sec & seeing, arcsec \\
 \hline
PA102    & 31.10/01.11.11 & VPHG1200B& 3700--5500 & 5.5  &4500 & 0.9 \\
PA102    & 01/02.11.11& VPHG1200R& 5750--7500 & 5.5  &4800 & 2.2--2.7 \\
PA187    & 28/29.03.11&VPHG940@600 & 3700--8500 & 6.5  &2700 & 2.2 \\
PA187    & 31.10/01.11.11& VPHG1200R& 3700--5300 & 5.5  &1800 & 0.9 \\
PA304    & 22/23.12.11& VPHG940@600& 3600--8400 & 6.5 & 3600 & 1.8--2.4 \\
PA347    & 22/23.12.11&VPHG940@600 & 3700--8500 & 6.5 & 5400 & 1.6--2.0 \\
\hline
\end{tabular}
\end{table*}

Radio observations of the galaxy \citep{synchro, braun, heald} revealed
continuum radio emission from the region of bright emission nebulae and, in particular,
from the brightest eastern chain. The polarised radio emission is also coincident with
this chain \citep{heald}. \citet{synchro} identified the synchrotron component of radio
emission in the eastern chain of bright nebulae, and this led them to suspect that
the region may contain supernova remnants.

The main goal of our paper is to perform a detailed study of the emission spectrum
and estimate the metallicity of ionised gas in the Ho~II galaxy by analysing the
spectroscopic observations made at the 6-m telescope of the SAO RAS. We also use the
available results of earlier spectroscopic observations of the galaxy made by \citet{lee}
and \citet{croxall} combined with \emph{HST} archival imaging data to estimate the metallicity
in the same way.

We used the seven currently most popular methods to estimate the oxygen abundance, which
determines the metallicity of the interstellar medium. Our aim was to compare different
techniques by applying them to the observations of a large number of \HII\ regions in
the Holmberg~II, IC~10, and VII~Zw~403 galaxies in order to choose
the optimal method for analysing the metallicity of gas in Irr galaxies.

We also attempted to detect the optical emission from the above mentioned hypothetical
supernova remnants using the $I($\SII$)/I({\rm H}\alpha)$ line intensity ratio in the
spectrum.

The next sections describe our observations, the data reduction technique, present the
results, their discussion, and the main conclusions of this paper.

We defer a comparison of the results of spectroscopic observations with the detailed
analysis of the kinematics of ionized and neutral gas and with IR emission of the dust
component of the galaxy to our forthcoming paper (Wiebe et~al., in preparation).

\section{Observations and data reduction}

The observations were  made at the prime focus of the SAO RAS 6-m telescope using
the SCORPIO multi-mode  focal reducer  \citep{scorpio1}  and its new version SCORPIO-2  \citep{scorpio2}.
When operated in the long-slit mode, both devices have the same slit size (6~arcmin $\times$  1~arcsec)
with a scale of 0.36~arcsec per pixel. However, with a  similar  spectral resolution \mbox{SCORPIO-2}
provides a twice larger spectral  range:   the  VPHG940@600 grism covers the  wavelength interval
spanning from  3700 to 8500~\AA, whereas with the SCORPIO we used two grisms VPHG1200B and VPHG1200R
for the green and red spectral regions, respectively. The CCDs employed were an \hbox{EEV 42-40} in the SCORPIO and \hbox{E2V 42-90} (with the sensitivity peak at redder wavelengths) in the SCORPIO-2.

Table~\ref{tab_observ} gives the log of observations: for each spectrum designated by
the position angle of the spectrograph slit it lists the observing date, grism employed,
spectral range $\Delta\lambda$, spectral resolution $\delta\lambda$ (estimated by the
FWHM of air glow lines), total exposure $T_{exp}$, and seeing.

Data reduction was performed in a standard way using the IDL software package
developed at the SAO RAS for reducing the long-slit spectroscopic data obtained with
the SCORPIO and SCORPIO-2.

To increase the signal-to-noise ratio for weak emission nebulae, we binned spectrograms
into 5--6 pixel bins along the slit prior to reduction. The binning size was chosen
to make the resulting spatial resolution for a single slit position consistent
with the maximum `seeing'. To compute the correction for the spectral sensitivity
of each grism and convert the spectra to the absolute intensity scale,
we observed the spectrophotometric standards Hz~44, Feige~34, and AGK+81d266
immediately after the object at a close zenith distance.

We used single- or two-component Gaussian fitting to measure the integrated
fluxes of emission lines. We estimated the errors of the measured fluxes by analysing
the synthetic spectra with the given signal-to-noise ratio. In this paper we use only
the spectral lines with the signal-to-noise
ratios greater than 3.

The error intervals listed in the tables and indicated in the
figures below correspond to the
$3\sigma$ confidence level.

\section{ Results of observations}

\begin{figure*}
\includegraphics[width=14cm]{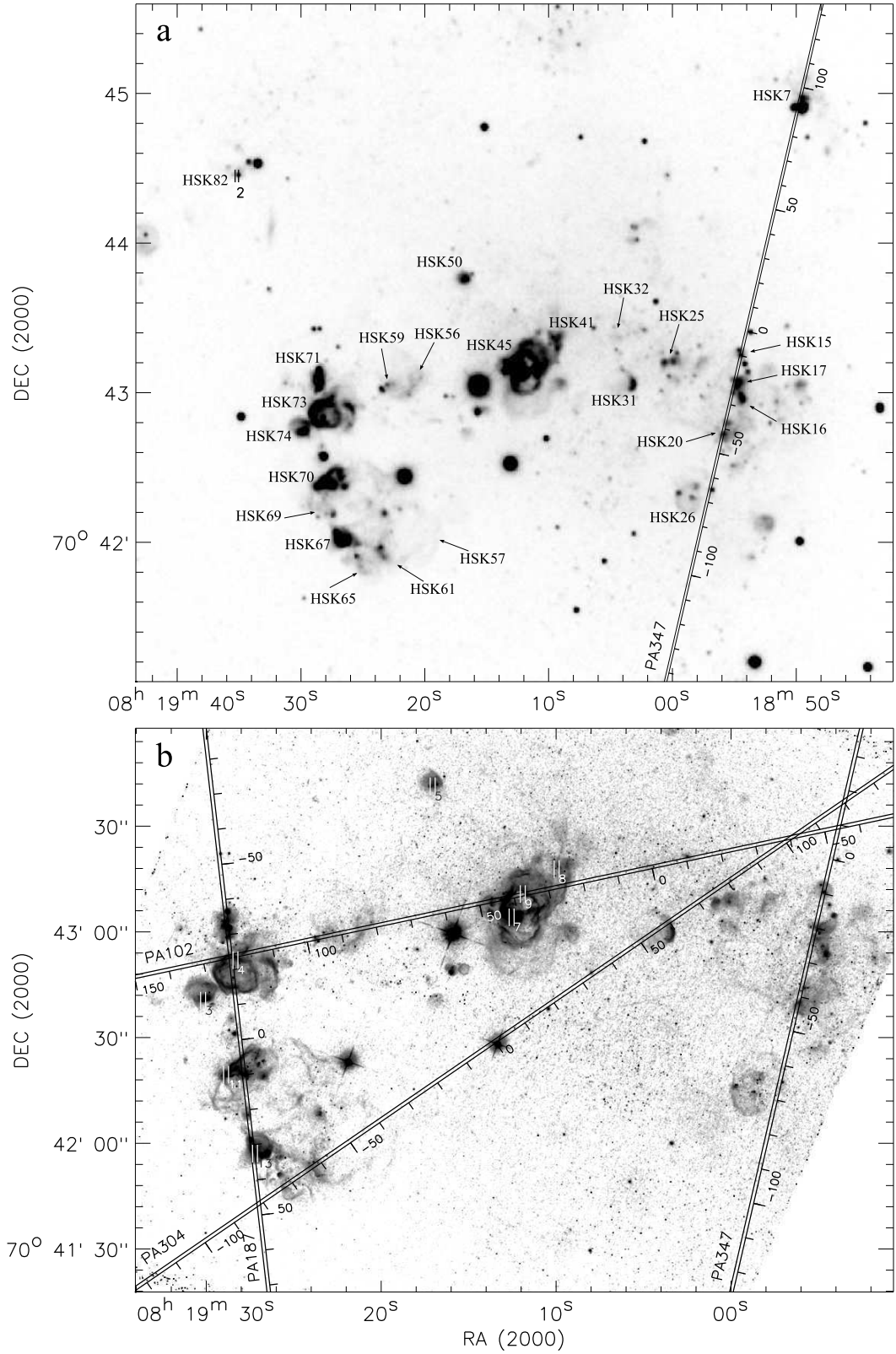}

\caption{The H$\alpha$+continuum images of the  Ho~II galaxy (panel `a' shows the
image taken with the 2.1-m  KPNO telescope and  panel `b', archival \emph{HST}
image (F658N/ACS WFC)). Panel `a' gives the designations of all \HII\ regions
discussed in this paper (denoted by the HSK numbers from the catalogue of \citealt{HSK}),
shows the position   of the spectrograph slit PA347 used in our observations and the
position of slit~\#2 in observations of \citet{croxall}. Panel `b' shows the
positions of the spectrograph slits in our observations; the short white line segments indicate
the locations of the \citet{croxall} spectrograms marked by their numbers in accordance with
the above paper.}\label{fig_local}
\end{figure*}

Fig.~\ref{fig_local} shows the location of six spectrograms taken at  four slit
positions, designated   PA102, PA187, PA304, and PA347 in accordance with their position
angles. The two panels of the figure show the H$\alpha$ images of the galaxy's
star-forming regions. The upper image was taken with the 2.1-m KPNO telescope within
the framework of the SINGS survey \citep{SINGS} and the lower image was taken by the \emph{HST}
(application number 10522) and adopted from the \emph{Hubble} Legacy
Archive\footnote{http://hla.stsci.edu/}.

In the north--south direction the PA187 spectrogram crosses the following \HII\ regions
(named according to the catalogue of \citealt{HSK}): HSK~71, HSK~73, HSK~70
(the nebula surrounding an ultraluminous X-ray source -- ULX Holmberg~II X-1), HSK~69, and HSK~67.
The PA102 slit crosses the \HII\
regions HSK~32, HSK~41, HSK~45, HSK~56, HSK~59, and HSK~73. The PA347 slit covers the western
chain of \HII\ regions: HSK~7, HSK~15, HSK~17, HSK~16, HSK~20, and HSK~26.
The PA304 spectrogram is least informative, because it crosses faint \HII\ regions
HSK~65, HSK~61, HSK~57, HSK~31, and HSK~25.

\begin{figure*}
\includegraphics{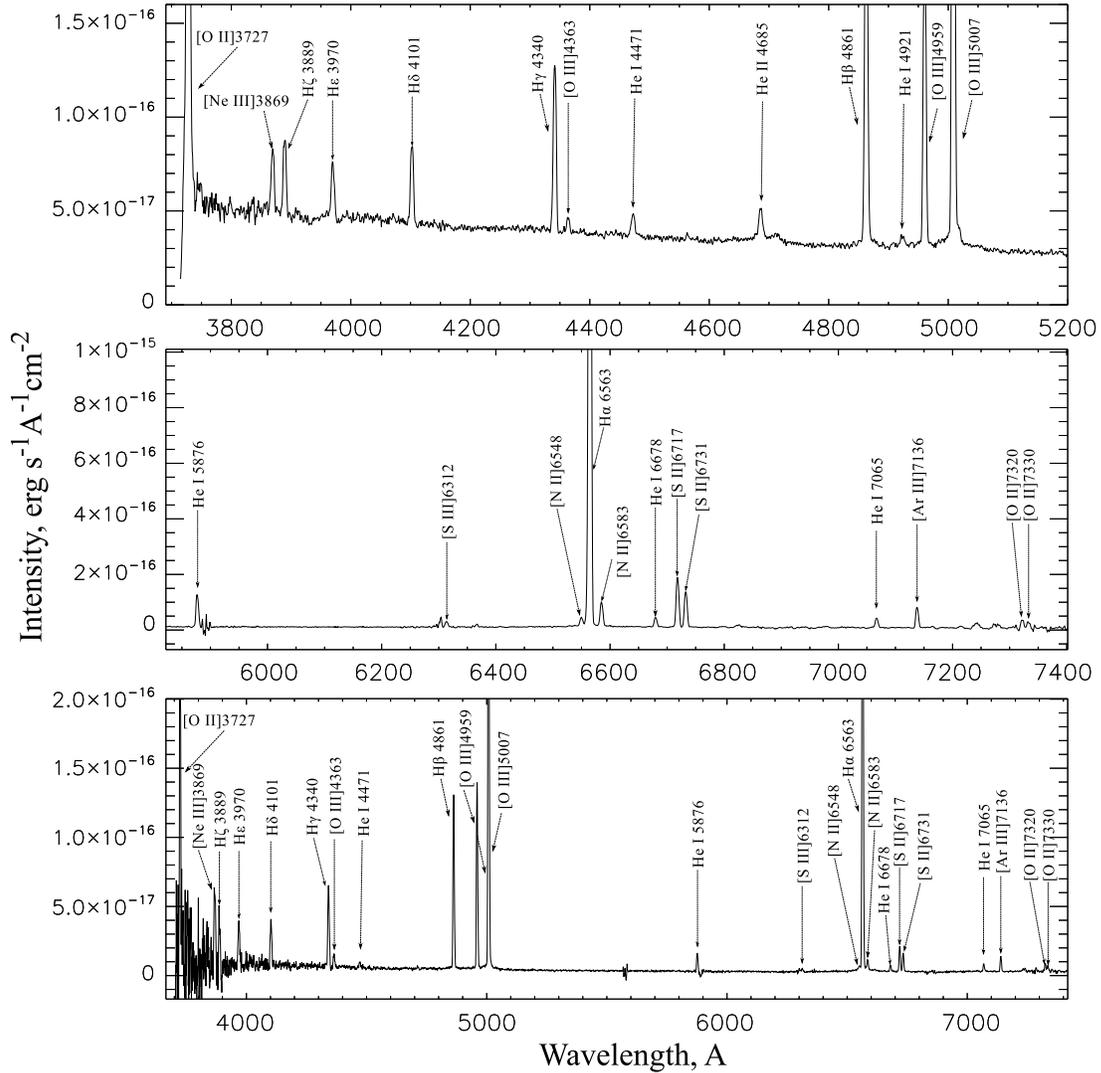}
\caption{Examples of spectra taken with different SCORPIO/SCORPIO-2 grisms. The top
panel shows the region between $-21.4$ and $-17.3$~arcsec along the  PA187 slit
(the central part of the HSK73 nebula, which exhibits bright emission in the \HeII\ $\lambda4686$ line)
observed with the VPHG1200B grism; the middle panel shows the region between  $40.7$ and $45.0$~arcsec
along the  PA102 slit (the HSK45 nebula) observed with the  VPHG1200R grism, and the bottom
panel shows the region between $94.6$ and $98.2$ along the PA347 slit (the HSK7 nebula)
observed with the VPHG940@600 grism.}\label{fig_spec_example}
\end{figure*}

Fig.~\ref{fig_spec_example} shows examples of spectrograms taken with different grisms.
The top panel shows the integrated spectrum of the part of the HSK73 \HII\ region crossed
by the PA187 slit (positions ranging from $-21.4$ to $-17.3$~arcsec along the slit) taken with the
VPHG1200B grism. It is the region where we found the brightest \HeII\ $\lambda$4686 emission line
(see section~\ref{sec:discussion}). The middle panel shows the integrated spectrum of the
eastern boundary of HSK45, which is the brightest nebula in the galaxy, taken with the VPHG1200R
grism (the PA102 spectrogram, positions ranging from $40.7$ to $45.0$~arcsec).
The bottom panel shows the integrated spectrum of the HSK7 region taken with the
VPHG940@600 grism (the PA347 spectrogram, positions ranging from $94.6$ to $98.2$~arcsec).
The spectrum of this region is of the highest quality among all the spectra
taken with this grism within the framework of this study (in the vicinity of the
\OII\ $\lambda3727$ and \OIII\ $\lambda4363$ lines its signal-to-noise ratio
is as high as 6--7).

Due to the wide spectral range of the VPHG940@600 grism there is possible second order contamination on the spectra obtained with this grism at wavelengths redder than 7000 \AA. But consideration of the continuum at this range (see, for example, bottom panel of the Fig.~\ref{fig_spec_example}) shows that possible systematic errors of emission line intensity measurements is no more than 12\% and lies in the uncertainties reported further.

To ensure the homogeneity of the inferred gas metallicities in the galaxy, we also use
the relative line intensities reported by \citet{lee} and \citet{croxall}. \citet{lee}
do not specify the locations of the spectra and only give the names of the \HII\
regions studied. Figure~\ref{fig_local} shows the locations of the spectra taken by
\citet{croxall} (we adopt the coordinates from the author-corrected astro-ph electronic
version of the paper). The spectrograms
are numbered in accordance with \citet{croxall}.

Table~2 lists our estimated
relative emission-line intensities in the individual \HII\ regions. We determined these
intensities by integrating the spectrum along the part of the nebula crossed by the
slit. The intensities are measured relative to  the H$\beta$ line intensity assuming that
$I({ \rm H\beta})=100$.

\medskip

We obtained two spectrograms for the PA187 slit. One of them, like in the case of PA304
and PA347, covers the entire spectral range from the blue to the red part, whereas the
other one contains only the blue part of the spectrum (see Table~\ref{tab_observ}). The
two spectrograms were taken on different nights with varying  seeing using different
instruments equipped with different CCDs. Hence a question naturally arises
whether the two spectra are on the same intensity scale. We compared the line
intensities in the blue part of the spectrum for each \HII\ region as estimated from both
spectrograms and found their intensity scales to be proportional to each other. Hence albeit
the line intensities measured by different spectrograms  differ from each other,
the ratios of the emission line intensities to the H$\beta$ flux are the same for
both spectra within the observational errors. That is why in the case of PA187
Table~2 lists relative line intensities measured either by
the `wide-range' spectrogram taken with the VPHG940@600 grism or
by the spectrogram taken with the VPHG1200B grism depending on the line wavelength.
The intensities measured by spectra taken with the `wide-range' VPHG940@600 grism are used for lines located on the red side
of \OIII\ $\lambda5007$, and those measured by the spectra taken with the VPHG1200B grism
are used for lines in the blue part of the spectrum, where the
sensitivity of the VPHG940@600 grism decreases.

The situation with PA102 is more complicated, because in this case two spectrograms have
been taken for the blue and red parts of the spectrum and their spectral ranges do not overlap,
preventing a similar comparison. Therefore to correctly estimate the ratios of line
intensities  on the red side of  \OIII\ $\lambda5007$ to H$\beta$,
we calculated the ratios of the corresponding line intensities to  H$\alpha$ and used
the `theoretical' $I({\rm H\alpha})/I({\rm H\beta})=2.809$  ratio for the average
electron temperature $T_e=13000$~K \citep{osterbrock}.

\medskip

The line intensity ratios listed in Table~2 are reddening corrected.
We determined the  $E(B-V)$ colour index from the Balmer decrement using the
theoretical intensity ratios  $I(\mathrm{H}\alpha)/I(\mathrm{H}\beta)=2.809$ and
$I(\mathrm{H}\gamma)/I(\mathrm{H}\beta)=0.4745$ for the electron temperature
$T_e=13000$~K \citep{osterbrock} and the extinction curve of \citet{cardelli}
as parametrised by \citet{fitz}. For the  PA102 slit we used only the
$I(\mathrm{H}\gamma)/I(\mathrm{H}\beta)$ ratio. The resulting $E(B-V)$
extinction values are listed in Table~3.

Our inferred colour excesses $E(B-V)$ are greater than the $E(B-V)=0.06 \pm 0.04$ estimate
reported by \citet{croxall}. We obtained a similar extinction value $E(B-V)=0.03$
towards Holmberg~II from the extinction map based on the data of the infrared sky survey
\citep{dustmap}.

The mean colour index averaged over all our spectrograms is $E(B-V)=0.17 \pm 0.08$.
It differs from the estimate of \citet{croxall} and this can be due to
the use of different extinction laws. Our inferred $E(B-V)$ value possibly corresponds to the stronger local
extinction, because all the bright \HII\ regions studied are located in the direction
of the maximum \HI\ column density in the `walls' of a giant cavity, and, most likely, are
partially embedded in this dense supershell.

\subsection{Estimation of the electron density and temperature.}

Our electron density $n_e$ and electron temperature $T_e$ estimates for a number of
\HII\ regions in Holmberg II have large uncertainties.

Table~3 lists the electron densities $n_e$ for each \HII\ region derived
from the \SII\ $\lambda6717/\lambda6731$ line intensity ratio. This line intensity
ratio proved to be very close to the limiting value for low densities and therefore we could determine
$n_e$ in a number of \HII\ regions only very approximately.

We determined the electron temperatures in \HII\ regions in terms of the so-called
two-zone model assuming that $T_e$ in the low and high ionization zones is the same
for all ions whose radiation emerges from these regions. Correspondingly, we assume
that the temperature in the high-ionization region (for the O$^{2+}$ and Ne$^{2+}$
ions) is equal to $T_e$(O~{\sc iii}), and the temperature in the low-ionization region
(for the O$^+$, N$^+$, and S$^+$ ions) is equal to $T_e$(O~{\sc ii}). We also adopt
the common extension of the model proposed by \citet{garnett}, which assumes that the
temperature inside the region of the S$^{2+}$ and Ar$^{2+}$ emission is equal to
the temperature $T_e$(S~{\sc iii}). 

We calculate the  $T_e$(O~{\sc iii}) temperature from the
\OIII\ ($\lambda4959$ + $\lambda5007$)/\OIII\ $\lambda4363$ line intensity
ratio \citep{osterbrock}. However, because of the low intensity of
the \OIII\ $\lambda4363$ line we could determine $T_e$(O~{\sc iii}) accurately enough
only for 10 \HII\ regions (including two different parts of the  HSK73 nebula) by the relation of \citet{ONS}:

\begin{equation*}
\label{eq_teo3}
  t=\frac{1.467}{\log Q_{3}-0.867-0.193\log t + 0.033t},
\end{equation*}
where $Q_{3}=I_\mathrm{[O\;III](\lambda4959+\lambda5007)}/I_\mathrm{[O\;III]\lambda4363}$ and $t=10^{-4}T_e(\mathrm{O}$~{\sc iii}).

\bigskip

We estimate the $T_e$(O~{\sc ii}) temperature in low-ionization regions by the
\OII\ ($\lambda3727 + \lambda3729$)/\OII\ ($\lambda7320 + \lambda7330$) line intensity
ratio using the following relation of \citet{pil_tem}:

\begin{equation*} \label{eq_teo2}
  t_2=\frac{0.96}{\log Q_{2}-0.86-0.38\log t_2 + 0.053t_2+\log(1+14.9x)},
\end{equation*}

where $x=10^{-4}n_et_2^{-1/2}$, $t_2=10^{-4}T_e$(O~{\sc ii}), and
$Q_{2}=I_\mathrm{[O\;II](\lambda3727+\lambda3729)}/I_\mathrm{[O\;II](\lambda7320+\lambda7330)}$.

\medskip

While determining $T_e$(O~{\sc ii}) we had to address a number of problems. First, in
observations of nearby objects, such as Holmberg~II, the \OII\ $\lambda7320+\lambda7330$
emission feature falls within the domain of strong atmospheric hydroxyl absorption lines,
which reduce the accuracy of the intensity measurements for these lines and hence
that of the inferred $T_e(\mathrm{O}$~{\sc ii}). Our failure to properly subtract the
contribution of the air glow lines in the spectra of some \HII\ regions makes
the corresponding \OII\ $\lambda7320+\lambda7330$ intensity estimates less reliable.

Second, the sensitivity of the CCD used in half of the observations
decreases sharply in the blue part of the spectrum. This prevented us from measuring
the \OII\ $\lambda3727+\lambda3729$ lines accurately enough in the PA304 and PA347 spectra,
making it impossible to calculate $T_e$(O~{\sc ii}). In the case of
the PA102 and PA187 spectra we derived the temperature in
the low-ionization regions only in four \HII\ regions, where the signal-to-noise ratio
in the \OII\ $\lambda7320+\lambda7330$ lines was greater than 4.

We   could thus `directly' determine the electron temperatures both in the low- and
high-ionization regions only in HSK45, HSK67, HSK71, and HSK73. However, even in these
nebulae the inferred  $T_e$(O~{\sc ii})  estimates are not accurate and most
likely represent the upper temperature limits for the corresponding regions. We therefore
used empirical methods to infer the temperatures in low-ionization regions in the cases
where we could determine $T_e$(O~{\sc iii}).

\medskip

Many methods are known for the empirical determination of electron temperature in
low-ionization regions from the known $T_e$(O~{\sc iii}) temperature. \citet{hagel}
compared some of the methods used to estimate the temperature in the  O$^+$ emission
region. \citet{perez_montero} pointed out that this temperature is strongly dependent
not only on $T_e$(O~{\sc iii}), but also on the electron density $n_e$. This imposes strong
constraints on the applicability of the empirical
$T_e$(O~{\sc ii}) on $T_e$(O~{\sc iii}) dependences. In particular, we did not use the dependence of  $T_e$(O~{\sc ii})
on $T_e$(O~{\sc iii}) found by \citet{perez_montero} for three electron density values, $n_e=10,
100$, and $500$ cm$^{-3}$, because of the above-mentioned low accuracy of the estimated $n_e$.

\citet{stas80, stas90} proposed a method, which gained widespread popularity.
\citet{izotov} proposed to determine the temperature in regions with ionization higher
and lower than in the  O$^{2+}$ emission region using a technique depending on the gas
metallicity. Furthermore, \citet{pil_tem} proposed an empirical formula for
estimating $T_e$(O~{\sc ii}) and $T_e$(N~{\sc ii}) from $T_e$(O~{\sc iii}). These authors
point out that the $T_e$(N~{\sc ii}) value should be preferred for the low-ionization regions
because of its lower dispersion. \citet{lopsan} proposed a calibration based on the model
spectra of \HII\ regions.

We used all the above mentioned methods to determine the temperatures in the
low-ionization regions. We found the calibration of \citet{lopsan} to be the most
consistent with the estimate determined from the spectra of four \HII\ regions, which
represents  an upper limit for the $T_e$(O~{\sc ii}) temperature in these nebulae. Hereafter
throughout this paper we adopt the estimates obtained using the above technique
obtained by the relation:

\begin{equation*}
\label{eq_teo2_lopsan}
T_e\mathrm{(O\;II)}=T_e\mathrm{(O\;III)}+450-70 \exp[(T_e\mathrm{(O\;III)}/5000)^{1.22}].
\end{equation*}

The temperatures $T_e$(O~{\sc ii}) estimated by other methods are
much higher than our inferred upper constraints for four \HII\ regions.

\medskip

The dependence of the temperature in the  S$^{2+}$ emission region on $T_e$(O~{\sc iii}) has a 
much lower scatter than the corresponding dependence for low-ionization regions
(see, e.g., \citealt{hagel}). We estimated $T_e$(S~{\sc iii}) using the equation
proposed by \citet{izotov} for different \HII\ region metallicities:

\begin{equation*}
\label{eq_tes3}
\begin{split}
t_3 & = -1.085+t\times (2.320-0.420t),{\rm\;for\;}12+\log\mathrm{(O/H)} \simeq 7.2 , \\
         & = -1.276+t\times (2.645-0.546t),{\rm\;for\;}12+\log\mathrm{(O/H)} \simeq 7.6, \\
         & = 2.967+t\times (-2.261+1.605t),{\rm\;for\;}12+\log\mathrm{(O/H)} \simeq 8.2, \\
\end{split}
\end{equation*}
where $t_3=10^{-4}T_e$(S~{\sc iii}),  $t=10^{-4}T_e$(O~{\sc iii}).
\bigskip

Table~3 lists the adopted electron temperatures in the regions of different
ionization.

\subsection{Estimation of the gas metallicity}

Currently, many methods are used to estimate the abundances of chemical elements
and primarily that of oxygen, which determines the metallicity of the interstellar
medium. The most popular is the so-called  `direct', or $T_e$ method, which allows
elemental abundances to be estimated from  the forbidden-line intensity ratios provided
that the electron temperature $T_e$ is known in the region where the corresponding
emission line forms.
However, we cannot always use this method since we have direct spectra-based
electron temperature estimates in the zones of different ionization  solely  for
four \HII\ regions. Only in 10 regions the temperatures in low-ionization zones
were estimated using empirical methods. We therefore used the $T_e$ method to estimate
the relative abundances of the O$^+$, O$^{2+}$, N$^+$, S$^+$, S$^{2+}$,
Ar$^{2+}$, and Ne$^{2+}$ ions in these regions wherever the signal-to-noise
ratio for the corresponding ion lines was greater than 3.

We estimated the relative ion abundances using  the relations from studies based on
modern atomic data: we used the equations from \citet{ONS}  to
compute the abundances of O$^+$ and O$^{2+}$ ions, and those from \citet{izotov}, to compute
the N$^+$, S$^+$, S$^{2+}$, Ar$^{2+}$, and Ne$^{2+}$ ionic abundances. For the unobserved
ionization stages we calculated the ionization correction factors ($ICF$) by
the relations adopted from \citet{izotov}, which allowed us to determine the abundances of
oxygen, sulphur, argon, and neon.

\medskip

Among the popular metallicity determination methods there are those based on the intensity ratios
of bright emission lines. These include the so-called `empirical' methods calibrated by \HII\
regions with bona fide oxygen abundance estimates and methods based on theoretical
photoionization models. In this paper we use six such methods:

\begin{figure*}
\includegraphics[width=15cm]{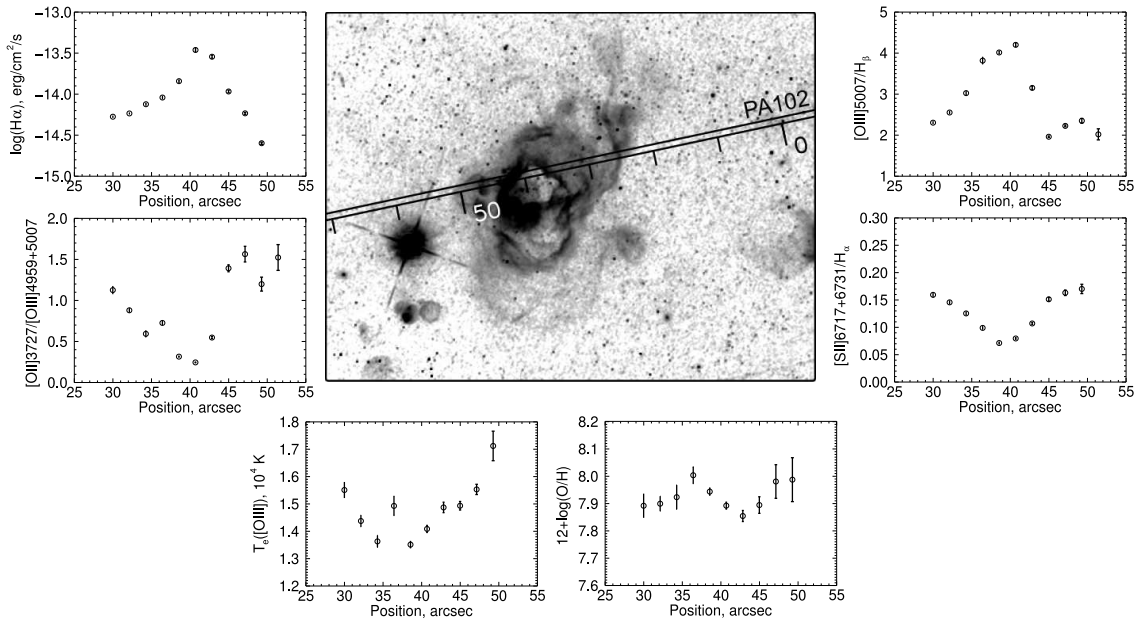}
\caption{The \HII\ region  HSK45. The central part of the figure shows the \emph{HST} H$\alpha$ image
together with the position of the PA102 slit.
The left, right, and bottom panels show the distribution of relative emission-line intensities,
electron temperature, and oxygen abundance, respectively, along the portion of the slit crossing
the HSK45 region.}\label{fig_hii_45_102}
\end{figure*}

\begin{enumerate}

\item The PT05 method \citep{PT05}, where the oxygen abundance
is fitted by a function of $R_2=I_{\mathrm{[O\;II]\lambda3727+\lambda3729}}/I_{\mathrm{H\beta}}$,  $R_3=I_{\mathrm{[O\;III]\lambda4959+\lambda5007}}/I_{\mathrm{H\beta}}$, $R_{23}=R_2+R_3$, and the excitation parameter
  $P=R_3/(R_3+R_2)$. It is one of the most widely used empirical methods with the applicability domain
confined to the $0.55 < P < 1$ interval;

 \item The ONS method \citep{ONS}, which allows the oxygen and nitrogen abundances to be determined as a function of parameters $R_2$,
 $R_3$,
 $N_2=I_{\mathrm{[N\;II]\lambda6548+\lambda6583}}/I_{\mathrm{H\beta}}$,
 $S_2=I_{\mathrm{[S\;II]\lambda6717+\lambda6731}}/I_{\mathrm{H\beta}}$, and $P$;

 \item The ON method \citep{ONS}, which is similar to the ONS method, but does not require the knowledge of $S_2$;

 \item The NS method \citep{NS}, which does not require
\OII\ $\lambda3727+\lambda3729$ intensity measurements and allows the oxygen and nitrogen abundances to be determined as functions of parameters $R_3$, $N_2$, and $S_2$;

\item The PP04 method \citep{PP04}, which allows the relative oxygen abundance to be determined from the parameter $O_3N_2=\log[(I_{\mathrm{[O\;III]\lambda5007}}/I_{\mathrm{H\beta}})/ (I_{\mathrm{[N\;II]\lambda6583}}/I_{\mathrm{H\alpha}})]$. This method is practically extinction
independent; it works in the $-1 < O_3N_2 < 1.9$ interval;

 \item The KK04 method (\citealt{KD02} with the new parametrisation by
\citealt{KK04}) based on theoretical photoionization models, which can be
used to determine the oxygen abundance as a function of parameter $R_{23}$ and
ionization parameter $q$, which, in turn, can be obtained from the $O_{32}=I_{\mathrm{[O\;III]\lambda4959+\lambda5007}}/I_{\mathrm{[O\;II]\lambda3727+\lambda3729}}$ line
intensity ratio.
\end{enumerate}

All the above methods can be used to determine the oxygen abundances,
$12+\log\mathrm{(O/H)}$  in the \HII\ regions. According to
the authors of original publications, these methods are accurate to about $0.1~\mathrm{dex}$
(the PP04 method is
less accurate, its error is of about $0.2-0.25~\mathrm{dex}$; the ON, NS, and ONS
methods have smaller errors -- of about $0.075~\mathrm{dex}$). The ON, NS, and ONS
methods can also be used to determine the nitrogen abundance and hence the
$\log\mathrm{(N/O)}$ abundance ratio, which is important for  chemical evolution
models of galaxies.

\medskip

The chemical abundances of the Holmberg~II \HII\ regions were earlier estimated by
\citet{masegosa}; \citet{lee}, and \citet{croxall}. \citet{moustakas} summarised these
results and reported the galaxy-averaged abundances and average abundances at different
galactocentric distances. We use all these data to compare with our results.

\citet{croxall} and \citet{lee} report relative line intensities, which we
used to determine the elemental abundances  by applying all the methods
employed in this paper.

Table~3 lists the resulting estimates of abundances of chemical elements and
individual ions   for the \HII\ regions based on our observations
and line intensities reported by \citet{croxall} and \citet{lee}.
Table~3 gives only the formal measurement errors, which do not include the uncertainties
of the each method itself.

Abundances obtained for similar slit positions based on our observations and those
reported in previous studies mostly agree well with each other within the quoted observational errors.
The oxygen abundances reported by \citet{croxall} and \citet{lee} slightly differ from
those calculated in this study using the fluxes reported by the above authors. This discrepancy
is due to the use of different atomic data. For some \HII\ regions, the  abundances determined in this paper
do not match those obtained by \citet{croxall} and \citet{lee}  because we
observed different parts of the nebulae, which show  significant abundance
variations (see Section~\ref{sec: abu_regs}).

\section{DISCUSSION}
\label{sec:discussion}

\subsection{Emission spectra of individual \HII\ regions}
\label{sec: abu_regs}

We analysed the variations of the emission-line intensity and physical conditions inside the 
\HII\ regions in the brightest nebulae HSK45, HSK73, HSK70, and HSK67. Note
that  the detailed distribution of metallicity along an individual
nebula could be determined only using the  `direct' method. Empirical methods can be applied only to the
entire nebula as a whole -- when they are used to determine the local chemical composition, the 
results are seriously affected by the emission stratification inside the region.

\medskip

\begin{figure*}
\includegraphics[width=15cm]{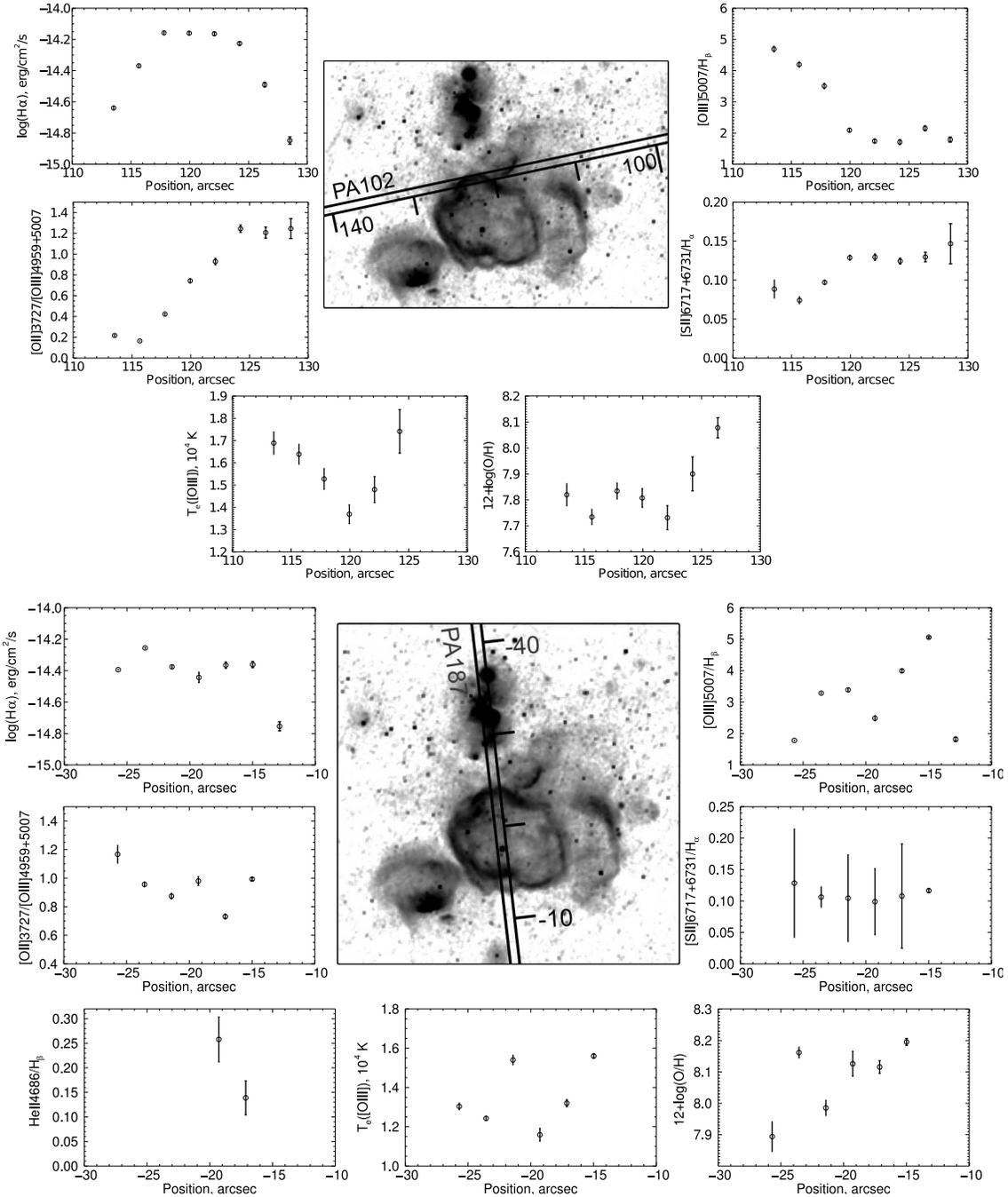}

\caption{The \HII\ region HSK73. The top and bottom panels show the results obtained
along the  PA102 and PA187 slits, respectively. The central part of each figure shows
the \emph{HST} H$\alpha$ image and the corresponding
slit position. Each figure shows  the distribution of relative emission-line
intensities, electron temperature, and oxygen abundance along the part of the slit
crossing this region.}\label{fig_hii_73}
\end{figure*}

{\bf HSK45.} This is the largest \HII\ complex seen in the H$\alpha$ images of the galaxy.
Its angular size is of about  25--30~arcsec, corresponding to the linear size of
410--490~pc. According to the estimates of \citet{stewart}, it is a young region with an
age of  2.5--3.5~Myr. Its composite multishell structure conspicuously shows up in the images.

Fig.~\ref{fig_hii_45_102} shows the image of the HSK45 \HII\ region and the
distribution of diagnostic line ratios along the part of the PA102 slit crossing the
nebula. The  \OIII\ $\lambda4959+\lambda5007$/H$\beta$ intensity ratio in the crossed
parts of the shells is equal to about 2 and reaches 4.5 in the faint central area. The
gas ionization degree at the centre is higher than at the edges of the complex. The
decrease of the \OIII\ $\lambda4959+\lambda5007$/H$\beta$ ratio at the periphery can be
explained by the fact that the oxygen emission there is dominated by O$^+$, and the
decrease of the \SII\ $\lambda6717+\lambda6731$/H$\alpha$ toward the centre of the
region could be caused by the high ionization stage of sulphur represented by S$^{2+}$ (possibly, in the vicinity
of an ionizing cluster).

\begin{figure*}
\includegraphics[width=15cm]{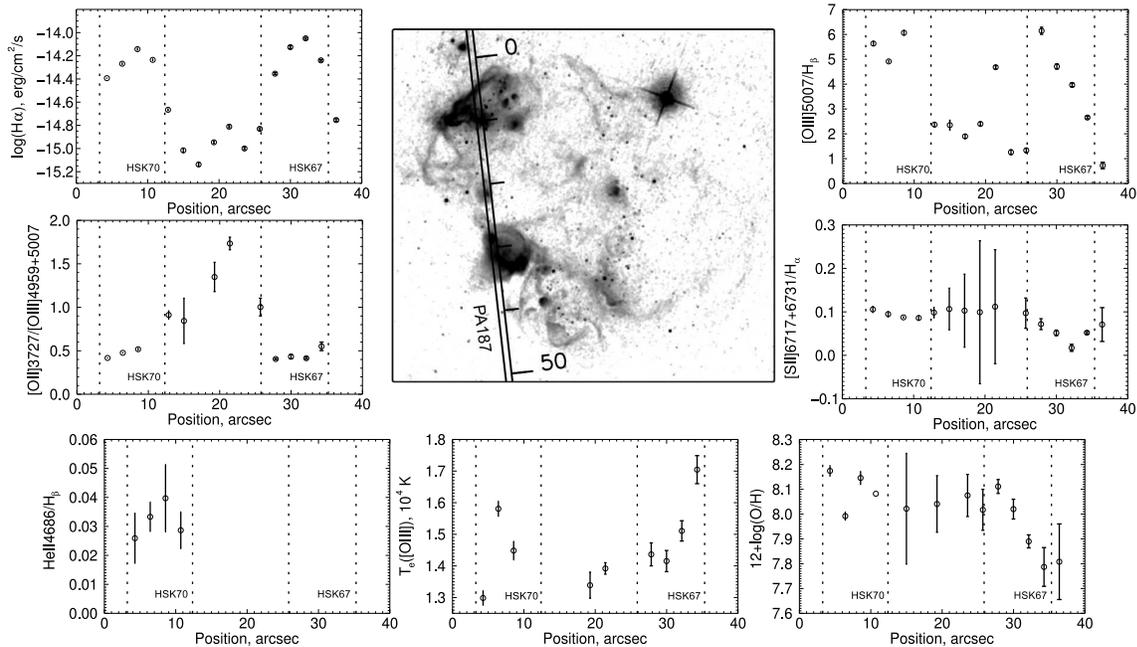}
\caption{The \HII\ regions HSK70 and HSK67. The central part of the figure shows the H$\alpha$
image (taken from the \emph{HST} archive). Also shown is the position of the  PA187 slit.
The other panels show the distribution of relative emission-line intensities, electron temperature,
and oxygen abundance along the slit.}\label{fig_hii_70_187}
\end{figure*}

The HSK45 region exhibits a non-uniform distribution of $T_e$ along the slit.
The relative oxygen abundance also varies along the region, however, these variations are
not so strong, and metallicity is more or less constant and consistent with
the region-averaged value listed in Table~3.

\medskip

{\bf HSK 73.} This region, like HSK45, is one of the brightest and most extended among those
seen in the H$\alpha$ image of the galaxy (see Fig.~\ref{fig_local}). The complex has an angular
size of about 15--20~arcsec  (linear size 250--330~pc). It is also a young region with
an age of about  $t=2.5-3.5$~Myr \citep{stewart}, characterised by a composite shell
structure. The centre of the region hosts the most massive stellar cluster of
those studied by \citet{cook} in the HoII galaxy.

Fig.~\ref{fig_hii_73} shows the image of the HSK73 region and the distribution of
relative emission-line intensities, the electron temperature and oxygen abundance along the
parts of the PA102 and PA187 slits crossing the complex from west to east and from
north to south, respectively.

The \OIII\ $\lambda4959+\lambda5007$/H$\beta$ intensity ratio is constant in the north-eastern part
of the region and increases significantly in its north-western part; this
ratio can  also be seen to increase from north to south. The
\OII\ $\lambda3727$/\OIII\ $\lambda4959+\lambda5007$ ratio, which characterises the degree of gas
ionization, decreases with increasing \OIII\ $\lambda4959+\lambda5007$/H$\beta$. This behaviour
indicates that gas ionization is higher in the north-western and southern parts of the
region and that the cause is photoionization.

We found a rather bright \HeII\ $\lambda4686$ line emission in the central part of the
region  (positions ranging from  $-17$ to $-20$~arcsec on the PA187 slit). Note that
the helium line emission region is compact and no such emission is present in the
spectrum of the outer shell. The \HeII\ line emission may supposedly be due to the fact
that here the slit
crosses a compact nebula ionised by a Wolf--Rayet star or by another source of intense
ultraviolet radiation.

The distribution of oxygen abundance along the portions of the PA102 and PA187 slits
crossing HSK73 demonstrates a constant gas metallicity in the north-western
outskirts of the region and a possible north-to-south metallicity gradient.
Oxygen abundance also increases sharply near the north-eastern
part of the shell.

\medskip

{\bf HSK 70 and HSK 67.} These two \HII\ regions are more compact than HSK45 and HSK73:
the size of HSK70 is of about 10--15~arcsec  (160--250~pc) and that of HSK67, about 8~arcsec (130~pc).
\citet{stewart} estimated the ages of these regions to be 3.5--4.5 and
2.5--3.5~Myr, respectively.

Fig.~\ref{fig_hii_70_187} shows the \emph{HST} H$\alpha$ images of  HSK70 and HSK67,
the position of the PA187 spectrograph slit,
and the distribution of relative line intensities, electron temperature, and
oxygen abundance along the slit.

The image of the extended surroundings of HSK70 and HSK67 reveals a faint shell-like
structure with a size of 40--50~arcsec (650--820~pc), which includes the two bright nebulae.

The only known ultraluminous X-ray source (ULX) in the galaxy, Holmberg~II X-1, is
located at the eastern boundary of the HSK70 region. This source completely determines
the kinematics and emission spectrum of the region (for a detailed study of the nebula
surrounding the source see \citealt{ulx}). The spectrograph slit in our
observations crosses the region 5~arcsec West of the source, but even at this distance
(80~pc) a strong \HeII\ $\lambda4686$ line emission of the nebula surrounding the ULX is
observed (see Fig.~\ref{fig_hii_70_187}).

The \OIII\ $\lambda4959+\lambda5007$/H$\beta$ intensity ratio is high inside the two
bright \HII\ regions and decreases in the space between them. The \OII\
$\lambda3727$/\OIII\ $\lambda4959+\lambda5007$ ratio, on the contrary, is low inside
the regions and  higher in the space between. This fact indicates that gas ionization inside
HSK70 and HSK67 is higher than in the outskirts of the two bright nebulae.

\begin{table*}
\caption{Oxygen abundances in the IC~10 and VII~Zw~403 galaxies estimated using different methods}\label{tab_met_ic_zw}
\begin{tabular}{clccccccc}
\hline
Galaxy & \HII\ region & \multicolumn{7}{c}{$12+\log({\rm O/H})$ estimated using different methods} \\

 & \&\ slit PA &$T_e$ & PT05 & ONS & ON & NS & PP04 & KK04 \\
\hline
IC~10 & HL111a PA0 & $8.28\pm0.04$ &$8.34\pm0.24$ &$8.14\pm0.06$ &$8.05\pm0.05$ &$8.16\pm0.02$ &$8.36\pm0.01$ &$8.27\pm0.28$ \\
IC~10 & HL111b PA0 & $8.29\pm0.04$ &$8.27\pm0.22$ &$8.18\pm0.06$ &$8.11\pm0.05$ &$8.22\pm0.03$ &$8.41\pm0.01$ &$8.23\pm0.29$ \\
IC~10 & HL111c PA0 &$8.24\pm0.02$ &$8.24\pm0.15$ &$8.03\pm0.03$ &$7.92\pm0.03$ &$8.07\pm0.01$ &$8.18\pm0.04$ &$8.38\pm0.11$ \\
IC~10 & HL111e PA0 & $8.24\pm0.07$ &$8.29\pm0.38$ &$8.12\pm0.12$ &$8.05\pm0.09$ &$8.15\pm0.04$ &$8.29\pm0.01$ &$8.28\pm0.45$ \\
IC~10 & HL111d PA268& $8.05\pm0.05$ &$8.28\pm0.11$ &$8.12\pm0.05$ &$8.07\pm0.04$ &$8.16\pm0.03$ &$8.26\pm0.05$ &$8.33\pm0.08$ \\
IC~10 & HL111c PA268& $8.15\pm0.03$ &$7.94\pm0.10$ &$7.86\pm0.03$ &$7.71\pm0.03$ &$7.70\pm0.03$ &$8.18\pm0.01$ &$8.32\pm0.02$ \\
IC~10 & HL111a PA331& $8.20\pm0.04$ &$8.05\pm0.34$ &$8.14\pm0.09$ &$8.17\pm0.07$ &$8.19\pm0.03$ &$8.13\pm0.01$ &$8.35\pm0.23$ \\
IC~10 & HL50 PA331&$8.19\pm0.20$ &$8.30\pm0.16$ &$8.07\pm0.34$ &$8.04\pm0.31$ &$8.01\pm0.13$ &$8.42\pm0.01$ &$-$ \\
IC~10 & HL45 PA45&$8.15\pm0.05$ &$7.92\pm0.17$ &$7.79\pm0.10$ &$7.84\pm0.08$ &$7.84\pm0.04$ &$8.31\pm0.01$ &$8.16\pm0.20$ \\
IC~10 & HL50 PA45&$8.25\pm0.02$ &$8.10\pm0.16$ &$8.24\pm0.03$ &$8.27\pm0.02$ &$8.29\pm0.01$ &$8.23\pm0.01$ &$8.35\pm0.10$ \\
VII~Zw~403 & \#1 & $7.71\pm0.01$ &$7.43\pm0.03$ &$7.72\pm0.02$ &$7.80\pm0.02$ &$7.77\pm0.01$ &$7.96\pm0.03$ &$7.90\pm0.18$ \\
VII~Zw~403 & \#3 & $7.73\pm0.01$ &$7.61\pm0.09$ &$7.68\pm0.02$ &$7.61\pm0.02$ &$7.62\pm0.01$ &$8.06\pm0.02$ &$8.02\pm0.19$ \\
VII~Zw~403 & \#4 & $7.75\pm0.02$ &$7.71\pm0.05$ &$7.71\pm0.03$ &$7.65\pm0.04$ &$7.67\pm0.03$ &$8.05\pm0.02$ &$8.09\pm0.14$ \\
\hline
\multicolumn{9}{p{.98\textwidth}}{Note. The \HII\ regions in IC~10 are designated in accordance with
the list of \citet{IC10_HL} and by the position angle of the spectrograph slit in our observations
of this galaxy \citep{ic10_spec, ic10_mpfs}.
The \HII\ regions of VII~Zw~403 are identified by their numbers used in our previous studies \citep{zw_1, zw_2}.}

\end{tabular}

\end{table*}

Gas metallicity in the HSK70 and in its outskirts is more or less the same and coincides
with the region-averaged value. The  HSK67 nebula exhibits a sharp decrease
of oxygen abundance in the north-south direction.

\medskip

The \SII\ $\lambda6717+\lambda6731$/H$\alpha$ line intensity ratio does not exceed 0.2
along all the regions described above, thereby supporting the photoionization mechanism of gas excitation.

\subsection{Comparison of different methods of oxygen abundance determination}

In recent years several authors performed detailed comparisons of different
methods of elemental abundance estimation in order to determine the bona fide
applicability domain of every particular method (see, e.g., \citealt{kewley08},
\citealt{lopsan} and references therein).

Thus the researchers have since long questioned even the correctness of the results
obtained by the `direct' $T_e$ method. This method may yield
underestimated abundances in the case of small-scale temperature fluctuations inside the \HII\
region \citep{piembert} or significant temperature variations along the nebula
\citep{stas78, stas05}. A similar effect can be observed if the nebula
studied is not a single Stroemgren zone, but a composite \HII\ region containing
several ionization sources\footnote{This is the case with all the \HII\ regions we
studied in the Holmberg~II galaxy.}.

\citet{NO_OH} showed that  the $T_e$-method can yield overestimated N/O abundance ratios
for composite nebulae, whereas the empirical ON and NS methods
analysed by the above authors are free from this drawback.

\citet{lopsan} used model spectra of \HII\ regions to perform a detailed comparison of
the currently most popular methods for estimating the chemical composition. The above
authors showed that the $T_e$ method somewhat overestimates the oxygen, neon, and argon
abundances compared to the input model values. They also showed that the
results of the ON, NS, and ONS methods agree excellently with those of the $T_e$ method
when applied to the spectra of real objects. However, they yield underestimated results
compared to the input values when applied to model spectra. Another conclusion of \citet{lopsan}
is that the KK04 method based on photoionization models yields the results that
are most consistent with the chemical composition of the stellar population. The
results of this method also agree very well with those based on recombination lines.
The abundances determined using the KK04 method are systematically greater than those
given by the $T_e$ method.

\begin{figure*}
\includegraphics[width=15cm]{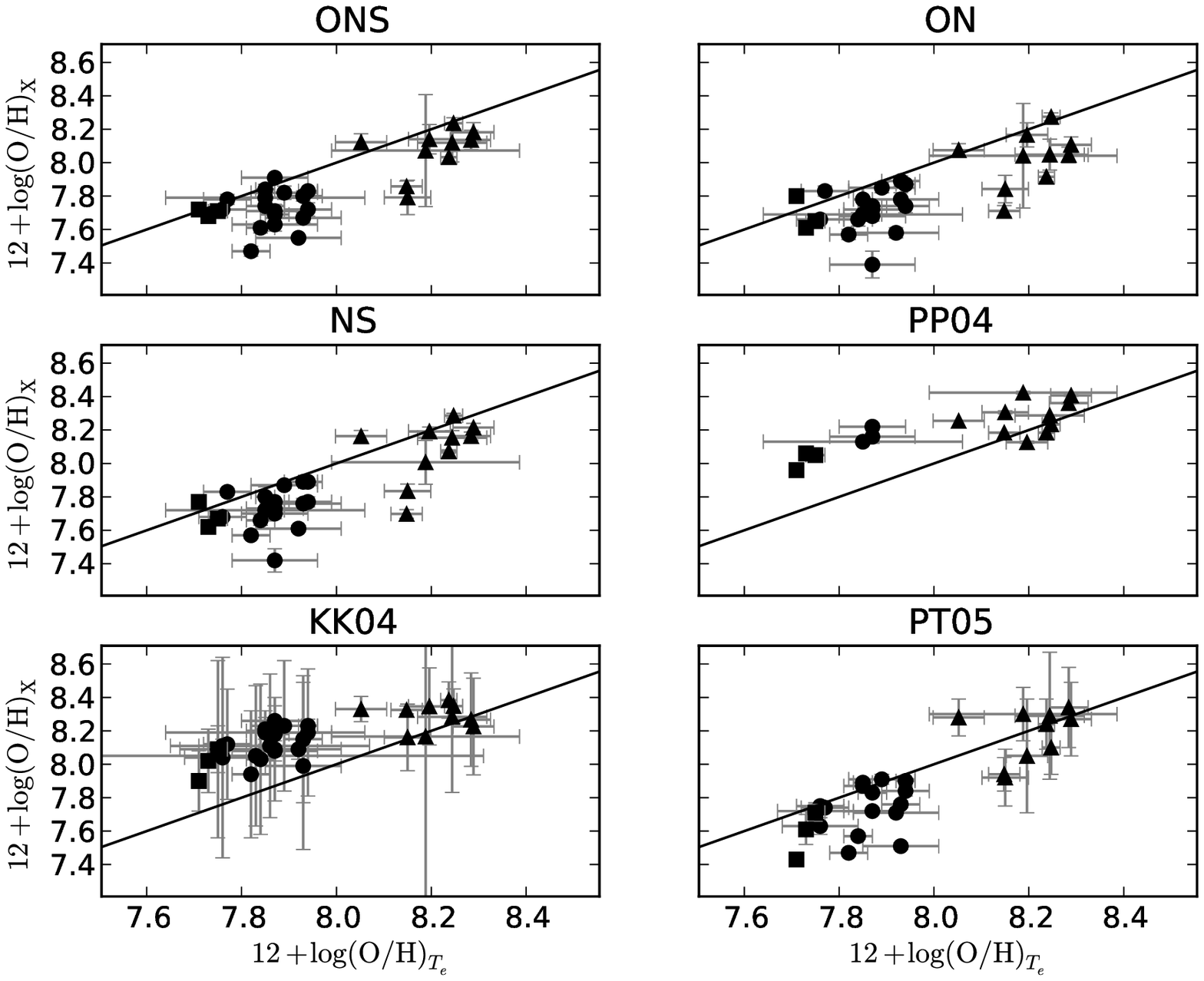}
\caption{Comparison of different methods for oxygen abundance estimation with the
`direct' $T_e$ method. The methods are designated by the following symbols:
PP04 -- \citet{PP04}; KK04 -- \citet{KK04}; PT05 -- \citet{PT05}; ONS and ON -- \citet{ONS}; NS -- \citet{NS}.
The circles, triangles, and squares show the results  of these methods applied 
to the  Holmberg~II, IC~10, and VII~Zw~403 galaxies, respectively.
The solid lines show the y=x relations for different methods.}\label{fig_met_compar}
\end{figure*}

\begin{figure*}
\includegraphics[width=15cm]{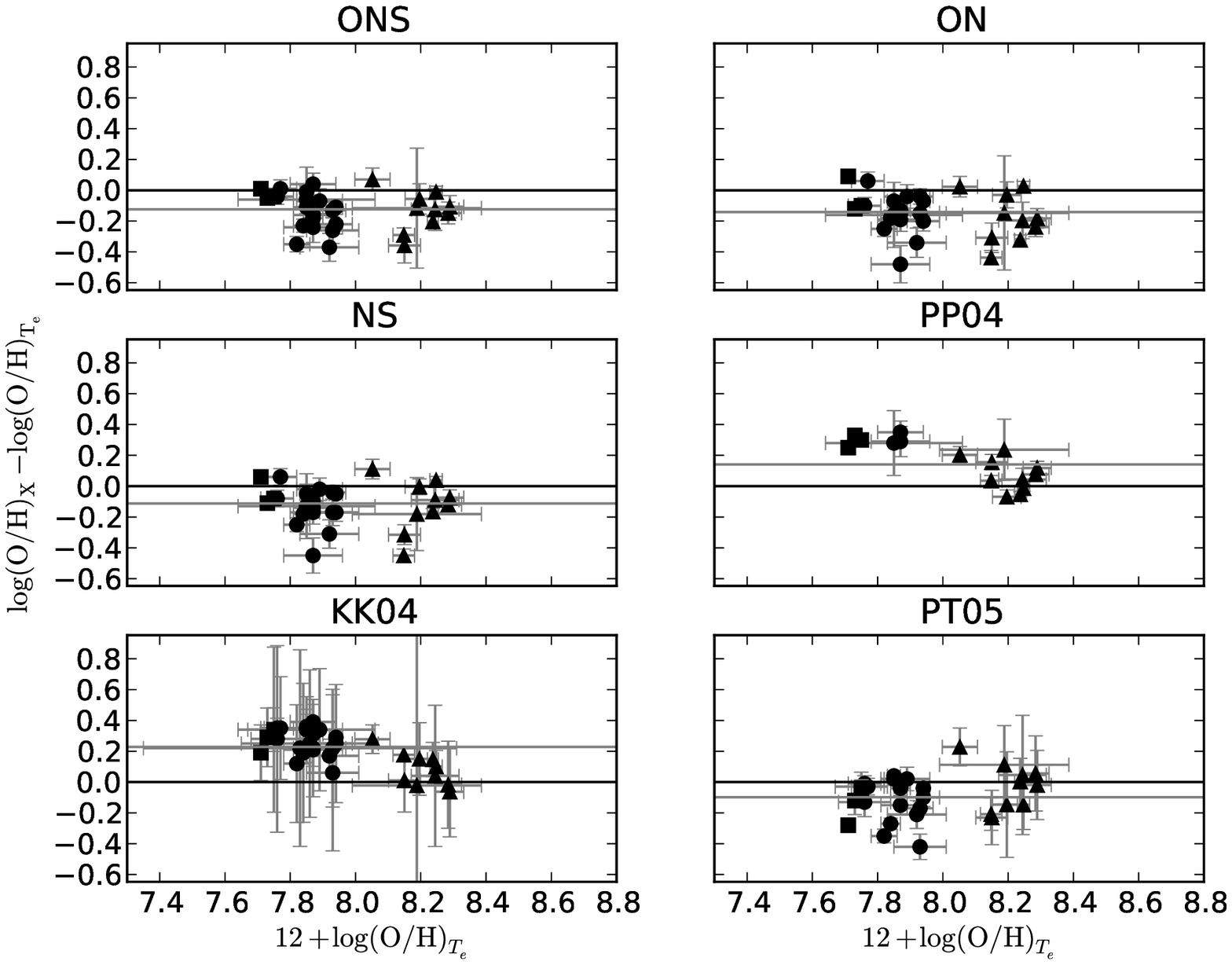}
\caption{Difference between the oxygen abundance estimates obtained using the `direct' $T_e$ method
and the six methods mentioned above. The methods employed are designated by the following symbols: PP04 -- \citet{PP04}; KK04 -- \citet{KK04}; PT05 -- \citet{PT05}; ONS and ON -- \citet{ONS}; NS -- \citet{NS}. The circles, triangles,
and squares show the results of the application of these methods to the
Holmberg~II, IC~10, and VII~Zw~403 galaxies, respectively.
The horizontal lines indicate the level of zero difference between the abundances given by the methods (black line) and the median deviation from the results of the $T_e$ method (grey line).}\label{fig_metdif_compar}
\end{figure*}

The Holmberg~II galaxy is an excellent test ground for comparing different methods of
estimating the chemical composition under the conditions of inhomogeneous and extended
\HII\ regions, where the effects like temperature variation are obviously present (see,
e.g., Figs.~\ref{fig_hii_45_102}, \ref{fig_hii_73}, \ref{fig_hii_70_187}). Indeed, most
of the bright \HII\ regions observed have angular sizes of  10~arcsec or more, implying
that their linear sizes exceed  160~pc, which is evidently greater than the size of a
classical Stroemgren zone around a single O-type star, and their ionization sources are
young star clusters. Furthermore, the H$\alpha$ images of the galaxy reveal the shell-like
structure of many of the \HII\ regions that we observed (see Fig.~\ref{fig_local}).

Our estimates of oxygen abundance in Holmberg~II \HII\ regions based on the $T_e$ method and six
empirical methods fall in two groups (see Table~3). The first group
contains the results  obtained using the $T_e$, ON, NS, ONS, and PT05 methods (the median
value $12+\log(\mathrm{O/H})=7.73\pm0.13$), and the second group those obtained using the PP04 and KK04 methods
(the median value $12+\log(\mathrm{O/H})=8.18\pm0.09$). \citet{lopsan} analysed
model spectra and showed that the PP04 method overestimates abundances at
$12+\log(\mathrm{O/H})<8.7$ and therefore the results based on this method may be incorrect.

\medskip

Our metallicity estimates obtained by applying different methods to the gas emission spectrum
in  Holmberg II  can also be compared with the results of our earlier study of the irregular
galaxy IC~10 \citep{ic10_spec, ic10_mpfs} and the BCD galaxy VII~Zw~403 \citep{zw_1, zw_2}.
Both galaxies were observed with the 6-m telescope of the SAO RAS using the SCORPIO focal reducer
operating in the long-slit spectrograph mode, and the MPFS multipupil fiber
spectrograph. We used these observations to measure the relative line intensities in 26 parts of 18 \HII\ regions in the IC~10 and in five extended \HII\ complexes  (giant \HII\ regions) in  VII~Zw~403.

In the above studies we estimated the gas metallicity using the PT05, PP04, and $T_e$
methods for the objects in IC~10 \citep{ic10_spec, ic10_mpfs}; and the ON, ONS, NS, and
$T_e$ methods for the objects in  VII~Zw~403 \citep{zw_1, zw_2}. To correctly compare the
three galaxies,  we re-estimate the oxygen abundances for IC~10 and VII~Zw~403 in this paper by applying all the methods that we used for Holmberg~II. 
Note that in original papers we made certain assumptions about $T_e$ in the \HII\ regions when estimating the abundances in IC~10 and VII~Zw~403 as described in \citep{ic10_spec} and \citep{zw_2} respectively. Our spectroscopic data allowed us to measure the intensity of the \OIII\ $\lambda4363$ line and hence the electron temperature only in one \HII\ region HL~111 in IC~10. We adopted
the $T_e$ measurements for another two nebulae (HL~45 and HL~50) from the literature. We obtained $T_e$ estimations from our spectra and from the literature for three \HII\ regions in VII~Zw~403. Further we use abundance measurements performed only for these regions with available $T_e$ estimations. The oxygen abundances for these regions of IC~10 and VII~Zw~403 obtained using all the methods considered are listed in Table~\ref{tab_met_ic_zw}.

We have thus analysed a total of 34 \HII\ regions in three galaxies using the `direct'
$T_e$ method and the six empirical and theoretical methods mentioned above,
and present the results in Fig.~\ref{fig_met_compar}.

Fig.~\ref{fig_metdif_compar} shows the differences between the results obtained using
these six methods as a function of the abundance determined using the direct $T_e$ method.

It follows from Table~3 and Figs.~\ref{fig_met_compar} and
\ref{fig_metdif_compar} that metallicities determined using different methods indeed
differ systematically, and the corresponding deviations agree well with the results of
simulated `model' computations by \citet{lopsan}. The estimates based on the empirical PT05, ON, NS, and ONS methods on the average agree well (although for some \HII\ regions, the ONS method yields overestimated abundances compared to the values determined using the other two methods). Note
that the PT05, ON, NS, and ONS methods somewhat underestimate the metallicities
compared to the $T_e$ method. The KK04 method, on the contrary,  systematically
overestimates metallicities compared to those, determined using the `direct' method. The PP04 method yields similar results. We applied  these methods to the
model data of \citet{lopsan} and found that the abundances determined using these
techniques agree well with the metallicity estimates based on recombination lines and
with the metallicity of the stellar population.

The observed variations of the estimated oxygen abundances of \HII\ regions obtained by different methods may be due to the complex shell-like and
clumpy structure of these regions. Indeed, it follows from Figs.~\ref{fig_hii_45_102}, \ref{fig_hii_73},
and \ref{fig_hii_70_187} that such regions exhibit appreciable, and in some cases quite important electron-temperature and
metallicity variations from the centre towards the periphery. When averaged over the entire region,
the results of the `direct' method
are more sensitive to the dense high-temperature `inclusions' than to the regions with the
`average' density and temperature; this may explain why the direct method yields
underestimated metallicities.

\subsection{Comparison of diagnostic diagrams with theoretical photoionization models.}

\begin{figure}

\includegraphics[width=8cm]{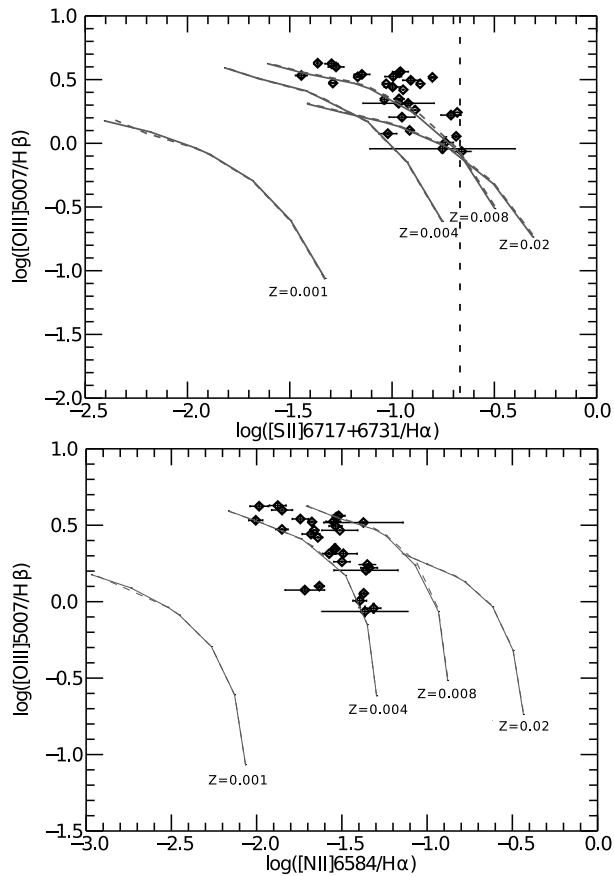}
\caption{Diagnostic diagrams based on the results of our observations of the
Holmberg~II galaxy and on the data of \citet{croxall}. Also shown are the theoretical
photoionization models by \citet{diagKewley} for different metallicities computed
assuming the continuous star formation over 5 Myr. The dashed line in the upper panel the ratio
$I($\SII\ $\lambda6717+\lambda6731)/I({\rm H}\alpha)=0.2$}\label{fig_diag_kew}
\end{figure}

Figure~\ref{fig_diag_kew} shows the diagnostic diagrams of relative line intensities ---
$I($\OIII\ $\lambda5007)/I({\rm H}\beta)$ --- as a function of $I($\NII\ $\lambda
6583)/I({\rm H}\alpha)$ or $I($\SII\ $\lambda6717+\lambda6731)/I({\rm H}\alpha)$. They
are based on our observations of the Holmberg~II galaxy and on the data of \citet{croxall}.
These diagrams are traditionally used to compare the observations with photoionization models.
The curves show the theoretical diagnostic diagrams of \citet{diagKewley} computed for
different metallicities assuming continuous star formation over 5 Myr.

\bigskip

It is evident from Fig.~\ref{fig_diag_kew} that the results of observations of \HII\
regions in the galaxy agree well with  models in the metallicity interval $Z=0.004 - 0.008$.
Given the current estimates of the solar metallicity, $Z_\odot=0.0134$ \citep{asplund},
this corresponds to $Z=(0.3 - 0.6)Z_\odot$. However, the mean oxygen abundance
in the galaxy estimated using the $T_e$, ON, NS, ONS and PT05 methods is $12+\log(\mathrm{O/H})=7.73 \pm 0.13$ ,
which yields $Z\simeq0.1Z_\odot$
(assuming $12+\log(\mathrm{O/H})_\odot=8.69$, see \citealt{asplund}). If we adopt
the mean oxygen abundance estimate $12+\log(\mathrm{O/H})=8.18 \pm 0.09$ given by
the PP04 and KK04 methods, the resulting metallicity becomes $Z\simeq0.3Z_\odot$ and
agrees well with photoionization models. This is not surprising given that both
methods have been calibrated by such models.

\bigskip

The \SII\ $\lambda6717+\lambda6731$/H$\alpha$ line intensity ratio is a commonly
used indicator of the shock excitation of gas. We tried to spectroscopically
confirm the presence of the shock excitation mechanism in the eastern chain of \HII\
regions, where \citet{synchro} identified the synchrotron component of the radio
emission, possibly indicative of the presence of supernova remnants. However, it is
evident from Fig.~\ref{fig_diag_kew} that the \SII\ $\lambda6717+\lambda6731$/H$\alpha$
line intensity ratio does not exceed  0.2 in all regions,
indicating the predominance of photoionization excitation in the areas considered.
It is not surprising that no evidence for the shock excitation of 
gas could be found in the optical spectra. At the relative distant galaxy the nebulae we analysed are
large complexes of  gas ionized by many stars, and optical emission of supernova remnants, if any,  is
hard to distinguish with the present spatial and spectral resolution. A detailed
investigation of the kinematics of the nebulae by Fabry-Perot interferometer observations
in H$\alpha$ and \SII\ lines that we defer to our forthcoming paper may be helpful
in case we find evidence of supersonic gas velocities.

\bigskip

\section{Conclusions.}

We report relative oxygen, nitrogen, sulphur, neon, and argon abundance estimates in the Holmberg~II galaxy based on the long-slit spectra taken at the 6-m telescope of the SAO RAS.

\medskip
The oxygen, nitrogen, sulphur, argon and neon abundances in Holmberg~II \HII\
regions that we report in this paper are the most detailed estimates
obtained for the galaxy used to date. According to our observations, the relative abundances of
these elements in individual \HII\ regions  (excluding regions with high uncertainties)
lie within the following intervals:
  $12+\log(\mathrm{O/H}) = 7.38 - 7.89$ (or $8.08 - 8.32$ depending on the method applied),   $12+\log(\mathrm{N/H}) = 5.92 - 6.51$,
   $12+\log(\mathrm{S/H}) = 6.01 - 6.20$, $12+\log(\mathrm{Ne/H}) = 6.89 - 7.07$, and $12+\log(\mathrm{Ar/H}) = 5.18 - 5.39$.
The average gas metallicity in the galaxy is $Z\simeq0.1\;Z_\odot$ (if computed using the $T_e$, PT05, ONS, ON or NS method)
or $Z\simeq0.3\;Z_\odot$ (if computed using the PP04 or KK04 method).

We used our observations, which cover most of the \HII\ regions in the galaxy, the data
reported by \citet{lee} and  \citet{croxall}, as well as the results of our earlier observations of
the Irr galaxy IC~10 and BCD galaxy VII~Zw~403  to compare the currently most popular
methods of the gas metallicity estimation in galaxies.

The results of this study, on the whole, confirm the conclusions of \citet{lopsan}
based on the analysis of `model' \HII\ regions. On the average, the estimates made
using the so-called PT05, ON, NS, and ONS methods agree well with each other (although the
ONS method yields overestimated abundances in some \HII\ regions, compared to the
results of other two methods). The PT05, ON, NS, and ONS methods yield lower metallicities
compared to the `direct' $T_e$ method. The PP04 and KK04 methods,
which were used in this paper, on the contrary, yield systematically overestimated
metallicities compared to the  $T_e$ method.

The PP04 method yields overestimated oxygen abundances for
$12+\log({\rm O/H}) < 8.7$ that are similar to those obtained by the KK04
method, leading us to conclude that the two methods yield less reliable estimates
for the objects considered than those given by the $T_e$, PT05, ON, NS and ONS methods.

We show the metallicity estimates based on the comparison of diagnostic diagrams with
the photoionization models to be less reliable than those determined from bright
oxygen lines.

\medskip

We tried to confirm the supernova remnants in the eastern chain of \HII\ regions
suspected by \citet{synchro}. However, we did not find any excessive
  $I($\SII\ $\lambda6717+\lambda6731)/I(\mathrm{H}\beta)$ line intensity ratio that would be
indicative of gas emission behind the shock.

\medskip

The distributions of relative line intensities along the slits crossing the brightest
and most extended \HII\ regions proved to be non-uniform from the viewpoint of the
structure and that of metallicity, temperature, and state of ionization.
This fact may play a key part in the determination of  metallicities of the entire
regions and cause discrepancies between the  estimates obtained using different
methods.

We found a bright \HeII\ $\lambda4686$ line emission in the central
part of the shell nebula HSK73. This emission may be indicative of the presence of a
Wolf--Rayet star or another energetic source of intense UV emission.

We will study the structure and kinematics of the gas and dust structure of the
Holmberg~II in our forthcoming paper (Wiebe et al., in preparation).

\bigskip

\section*{Acknowledgments}

This work was supported by the Russian Foundation for Basic Research (projects no.~10-02-00091 and 12-02-31356). O.V. Egorov
acknowledges the support from the Dynasty Foundation of Non-commercial Programs and the Federal Target
Program `Research and Pedagogical Cadre for Innovative Russia'
(state contract no.~14.740.11.0800). This research was partly supported by the RAS program of the fundamental investigation OFN-17 `Active processes in galactic and extragalactic objects'.

We are grateful to Kevin Croxall for verifying and sharing the corrected spectrograph slit
coordinates in the observations performed by his team, which we used in this work.
Also we thank Vera Arkhipova for helpful  discussions and Victor Afanasiev for his great contribution in the development of the spectroscopic technique at the 6-m telescope. We thank the anonymous referee for the constructive comments that improved the quality of the paper.

This work is based on the observations obtained with the 6-m telescope of the Special Astrophysical Observatory of the Russian Academy of
Sciences. The observations were carried out with the financial support of the Ministry of Education and Science of Russian Federation
(contracts no. 16.518.11.7073 and 14.518.11.7070).
This research has made use of the NASA/IPAC Extragalactic Database (NED) which is operated
by the Jet Propulsion Laboratory, California Institute of Technology, under contract with
the National Aeronautics and Space Administration (USA).

\label{lastpage}


\begin{thebibliography}{99}


\bibitem[\protect\citeauthoryear{Afanasiev \& Moiseev}{2005}]{scorpio1}
Afanasiev V.L., Moiseev, A.V., 2005 Astron. Letters, 31, 194

\bibitem[\protect\citeauthoryear{Afanasiev \& Moiseev}{2011}]{scorpio2}
Afanasiev V.L., Moiseev A.V., 2011, Baltic Astronomy, 20, 363

\bibitem[\protect\citeauthoryear{Arkhipova et~al.}{2007}]{zw_1}
Arkhipova V.P., Lozinskaya T.A., Moiseev A.V., Egorov O.V., 2007, Astron. Rep., 51, 871

\bibitem[\protect\citeauthoryear{Arkhipova et~al.}{2011}]{ic10_mpfs}
Arkhipova V.P., Egorov O.V., Lozinskaya T.A., Moiseev A.V., 2011, Astron. Lett., 37, 65

\bibitem[\protect\citeauthoryear{Asplund et~al.}{2009}]{asplund}
Asplund M., Grevesse N., Sauval A.J., Scott P., 2009, ARA\&{A} 47, 481

\bibitem[\protect\citeauthoryear{Bagetakos et~al.}{2011}]{bagetakos}
Bagetakos I., Brinks E., Walter F., de Blok W.J.G., Usero A., Leroy A.K., Rich J.W., Kennicutt R.C.Jr., 2011, AJ, 141, 23

\bibitem[\protect\citeauthoryear{Braun et~al.}{2007}]{braun}
Braun R., Oosterloo T. A., Morganti R., Klein U., Beck R., 2007, A\&A, 461, 455


\bibitem[\protect\citeauthoryear{Cardelli, Clayton \& Mathis}{Cardelli et~al.}{1989}]{cardelli}
Cardelli, J.A., Clayton, G.C., Mathis, J.S., 1989, ApJ, 345, 245

\bibitem[\protect\citeauthoryear{Cook et~al.}{2012}]{cook}
Cook D.O. et al., 2012, ApJ, 751, 100


\bibitem[\protect\citeauthoryear{Croxall et~al.}{2009}]{croxall}
Croxall K.V., van Zee L., Lee H., Skillman E.D., Lee J.C., C\^ot\'e S., Kennicutt R.C.,
Miller B.W., 2009, ApJ, 705, 723


\bibitem[\protect\citeauthoryear{Egorov \& Lozinskaya}{2011}]{zw_2}
Egorov O.V., Lozinskaya T.A., 2011, Ast. Bull., 66, 293

\bibitem[\protect\citeauthoryear{Fitzpatrick}{1999}]{fitz}
Fitzpatrick E.L., 1999, PASP, 111, 63

\bibitem[\protect\citeauthoryear{Garnett}{1992}]{garnett}
Garnett D.R., 1992, AJ, 103, 1330

\bibitem[\protect\citeauthoryear{H\"{a}gele et~al.}{2008}]{hagel}
H\"{a}gele G.F., D\'{\i}az \'A.I., Terlevich E., Terlevich R., P\'erez-Montero E., Cardaci M.V., 2008, MNRAS, 383, 209

\bibitem[\protect\citeauthoryear{Heald, Braun \& Edmonds}{Heald et~al.}{2009}]{heald}
Heald G., Braun R., Edmonds R., 2009, A\&A, 503, 409

\bibitem[\protect\citeauthoryear{Hodge \& Lee}{1990}]{IC10_HL}
Hodge P., Lee M.G., 1990, PASP, 102, 26

\bibitem[\protect\citeauthoryear{Hodge, Strobel \& Kennicutt}{Hodge et~al.}{1994}]{HSK}
Hodge P., Strobel N.V., Kennicutt R.C., 1994, PASP, 106, 309

\bibitem[\protect\citeauthoryear{Hunter, Hawley \& Gallagher}{Hunter et~al.}{1993}]{hunter}
Hunter D.A., Hawley W.N., Gallagher J.S., 1993, AJ, 106, 1797

\bibitem[\protect\citeauthoryear{Izotov et~al.}{2006}]{izotov}
Izotov Y.I., Stasi\'{n}ska G., Meynet G., Guseva N.G., Thuan T.X., 2006, A\&{A}, 448, 955

\bibitem[\protect\citeauthoryear{Karachentsev et~al.}{2003}]{karach03}
Karachentsev I.D. et~al., 2003, A\&{A}, 398, 479

\bibitem[\protect\citeauthoryear{Karachentsev \& Kaisin}{2007}]{karach07}
Karachentsev I.D.,  Kaisin S.S., 2007, AJ, 133, 1883

\bibitem[\protect\citeauthoryear{Kennicutt et~al.}{2003}]{SINGS}
Kennicutt R.C.Jr. et al., 2003, PASP, 115, 928

\bibitem[\protect\citeauthoryear{Kewley \& Dopita}{2002}]{KD02}
Kewley L.J., Dopita M.A., 2002, ApJS, 142, 35

\bibitem[\protect\citeauthoryear{Kewley \& Ellison}{2008}]{kewley08}
Kewley L.J., Ellison S.L., 2008, ApJ, 681, 1183

\bibitem[\protect\citeauthoryear{Kobulnicky \& Kewley}{2004}]{KK04}
Kobulnicky H.A., Kewley L.J., 2004, ApJ, 617, 240

\bibitem[\protect\citeauthoryear{Lee et~al.}{2003}]{lee}
Lee H., McCall M. L., Kingsburgh R. L., Ross R., Stevenson C.C., 2003, AJ, 125, 146

\bibitem[\protect\citeauthoryear{Lehmann et~al.}{2005}]{ulx}
Lehmann I. et al., 2005, A\&A, 431, 847

\bibitem[\protect\citeauthoryear{Levesque, Kewley \& Larson}{Kewley et~al.}{2010}]{diagKewley}
Levesque E.M., Kewley L.J., Larson K.L., 2010, AJ, 139, 712

\bibitem[\protect\citeauthoryear{L\'{o}pez-S\'{a}nchez et~al.}{2012}]{lopsan}
L\'{o}pez-S\'{a}nchez \'{A}.R., Dopita M.A., Kewley L.J., Zahid H.J., Nicholls D.C., Scharw\"{a}chter J., 2012, MNRAS, in press, astro-ph/1203.5021

\bibitem[\protect\citeauthoryear{Lozinskaya et~al.}{2009}]{ic10_spec}
Lozinskaya T.A., Egorov O.V., Moiseev A.V., Bizyaev D.V., 2009, Astron. Lett., 35, 730


\bibitem[\protect\citeauthoryear{Masegosa, Moles \& del Olmo}{Masegosa et~al.}{1991}]{masegosa}
Masegosa J., Moles M., del Olmo A., 1991, A\&A, 249, 505


\bibitem[\protect\citeauthoryear{Moustakas et~al.}{2010}]{moustakas}
Moustakas J., Kennicutt  R.C., Tremonti C.A., Dale D.A., Smith J.-D.T., Calzetti D., 2010, ApJS, 190, 233


\bibitem[\protect\citeauthoryear{Osterbrock \& Ferland}{2006}]{osterbrock}
Osterbrock D.E., Ferland G.J., 2006, Astrophysics of Gaseous Nebulae and Active Galactic Nuclei (2nd ed.). Univ. Science Book, Sausalito, CA

\bibitem[\protect\citeauthoryear{P\'{e}rez-Montero \& D\'{\i}az}{2003}]{perez_montero}
P\'{e}rez-Montero E., D\'{\i}az \'{A}.I., 2003, MNRAS, 346, 105

\bibitem[\protect\citeauthoryear{Pettini \& Pagel}{2004}]{PP04}
Pettini M., Pagel B., 2004, MNRAS, 348, L59

\bibitem[\protect\citeauthoryear{Piembert}{1967}]{piembert}
Piembert M., 1967, ApJ, 150, 825

\bibitem[\protect\citeauthoryear{Pilyugin \& Thuan}{2005}]{PT05}
Pilyugin L.S., Thuan T.X., 2005, ApJ, 631, 231

\bibitem[\protect\citeauthoryear{Pilyugin et~al.}{2009}]{pil_tem}
Pilyugin L.S., Mattsson L., V\'ilchez J.M., Cedr\'es B., 2009, MNRAS, 398, 485

\bibitem[\protect\citeauthoryear{Pilyugin, V\'{i}lchez \& Thuan}{Pilyugin et~al.}{2010}]{ONS}
Pilyugin L.S., V\'ilchez J.M., Thuan T.X., 2010, ApJ, 720, 1738

\bibitem[\protect\citeauthoryear{Pilyugin et~al.}{2012}]{NO_OH}
Pilyugin L.S., V\'ilchez J.M., Mattsson L., Thuan T.X., 2012, MNRAS, 421, 1624

\bibitem[\protect\citeauthoryear{Pilyugin \& Mattsson}{2011}]{NS}
Pilyugin L.S., Mattsson L., 2011, MNRAS, 412, 1145

\bibitem[\protect\citeauthoryear{Schlegel, Finkbeiner \& Davis}{Schlegel et~al.}{1998}]{dustmap}
Schlegel D.J., Finkbeiner D.P., Davis M., 1998, ApJ, 500, 525

\bibitem[\protect\citeauthoryear{Stasi\'{n}ska}{1978}]{stas78}
Stasi\'{n}ska G., 1978, A\&A, 66, 257

\bibitem[\protect\citeauthoryear{Stasi\'{n}ska}{1980}]{stas80}
Stasi\'{n}ska G., 1980, A\&{A}, 84, 320

\bibitem[\protect\citeauthoryear{Stasi\'{n}ska}{1990}]{stas90}
Stasi\'{n}ska G., 1990, A\&{A}S, 83, 501

\bibitem[\protect\citeauthoryear{Stasi\'{n}ska}{2005}]{stas05}
Stasi\'{n}ska G., 2005, A\&A, 434, 507

\bibitem[\protect\citeauthoryear{Stewart et al.}{2000}]{stewart}
Stewart S.G. et al., 2000, ApJ, 529, 201

\bibitem[\protect\citeauthoryear{Tonque \& Westphal}{1995}]{synchro}
Tongue T.D., Westphal D.J., 1995, AJ, 109, 2462

\bibitem[\protect\citeauthoryear{Walter et~al.}{2007}]{walter}
Walter F. et al., 2007, ApJ, 661, 102

\bibitem[\protect\citeauthoryear{Weisz et~al.}{2009}]{weisz}
Weisz D.R., Skillman E.D., Cannon J.M., Dolphin A.E., Kennicutt R.C.Jr., Lee J., Walter F., 2009, ApJ, 704, 1538


\end{thebibliography}
\end{document}